\documentclass[dvips,12pt]{article}
\usepackage{a4,graphics,psfig,t1enc}
\usepackage{epsf,amsmath,amssymb,latexsym}
\DeclareMathOperator{\rpart}{Re}
\DeclareMathOperator{\impart}{Im}

\begin{document}

\setcounter{page}{1}

{\small{\date{April 3, 2002}}}
\title{Orbit bifurcations and  wavefunction autocorrelations\footnote{Short
 title: Bifurcations and autocorrelations}}
\author{A.~B\"acker$^{1}$, J.~P.~Keating$^{2}$ and S.~D.~Prado$^{3}$ \\
{\small $1$ Abteilung Theoretische Physik, Universit\"at Ulm,} \\
{\small Albert-Einstein-Allee 11, D-89081 Ulm, Germany.} \\
{\small $2$ School of Mathematics, University of Bristol, Bristol BS8 1TW, UK.} \\
{\small $3$ Instituto de F{\'\i}sica, Universidade Federal do Rio
 Grande do Sul} \\
{\small P.O. Box 15051, 91501-970 Porto Alegre, RS, Brazil.}
}
\maketitle
\begin{abstract}

It was recently shown (Keating \& Prado, {\it Proc. R. Soc. Lond.
A} {\bf 457}, 1855-1872 (2001)) that, in the semiclassical limit,
the scarring of quantum eigenfunctions by classical periodic orbits 
in chaotic systems may be
dramatically enhanced when the orbits in question undergo
bifurcation. Specifically, a bifurcating orbit gives rise to a
scar with an amplitude that scales as $\hbar^{\alpha}$ and a
width that scales as $\hbar^{\omega}$, where $\alpha$ and $\omega$
are bifurcation-dependent scar exponents whose values are
typically smaller than those ($\alpha=\omega=1/2$) associated with
isolated and unstable periodic orbits. We here analyze the
influence of bifurcations on the autocorrelation function of
quantum eigenstates, averaged with respect to energy. It is shown
that the length-scale of the correlations around a bifurcating orbit
scales semiclassically as $\hbar^{1-\alpha}$, where $\alpha$ is the corresponding
scar amplitude exponent. This imprint of bifurcations on quantum
autocorrelations is illustrated by numerical computations for a
family of perturbed cat maps.
\end{abstract}

\newpage

\section{Introduction}

\hspace{\parindent}

The local statistical properties of quantum eigenfunctions
$\psi_n({\bf q})$ in classically chaotic systems are often
modelled in the semiclassical limit by treating the
eigenfunctions as random superpositions of plane waves (Berry
1977).  This allows, for example, for value distributions and
autocorrelation functions to be computed.  However, it ignores
contributions from the scarring of the eigenfunctions by
classical periodic orbits (Heller 1984, Kaplan 1999).
Understanding these contributions has been one of the central
goals in quantum chaology.  Our purpose here is to study the
influence of bifurcating periodic orbits on the autocorrelation
of eigenfunctions.

The semiclassical theory describing the scarring of wavefunctions
by isolated and unstable periodic orbits was developed by
Bogomolny (1988), who derived an explicit formula for the
contribution from a given such orbit to $|\psi_n({\bf q})|^2$,
averaged locally with respect to position and energy.  Berry
(1989) extended this approach to the Wigner functions associated
with energy eigenstates, again averaged locally with respect to
the energy (position averaging is unnecessary in this case).
Berry's formula also describes the influence of isolated and
unstable periodic orbits on the autocorrelation of the
eigenfunctions, because the Wigner function and the autocorrelation
function $\langle\psi_n({\bf q+L})\psi_n^*({\bf q})\rangle$ are
fourier transforms of each other (explicit formulae my be found in
Li \& Rouben 2001).  It was shown in Keating (1991) that
Bogomolny's and Berry's formulae are exact, rather than
semiclassical approximations, for quantum cat maps.

The extension of Bogomolny's theory to bifurcating periodic
orbits was made in Keating \& Prado (2001).  There it was shown
that orbits undergoing bifurcation give rise to scars in
$|\psi_n({\bf q})|^2$ (averaged locally with respect to position
and energy) with a width that scales as $\hbar^\omega$ and an amplitude
that scales as $\hbar^{\alpha}$ when $\hbar \rightarrow 0$, where
$\omega$ and $\alpha$ are bifurcation dependent scar exponents whose
values are positive and less than or equal to those ($\omega=\alpha=1/2$)
corresponding to isolated and unstable orbits.  For most of the
generic bifurcations $\omega<1/2$ and $\alpha < 1/2$, and so
bifurcations may be said to give rise to {\em superscars}.  This
mirrors the dominant influence they have previously been shown to
exert on spectral fluctuation statistics (Berry {\em et al.}
1998, Berry {\em et al.} 2000).

The question we address here is how bifurcations influence the
autocorrelation of the eigenfunctions. Our main result is that
the length-scale of the correlations around a bifurcating orbit
scale as $\hbar^{1-\alpha}$ when $\hbar \rightarrow 0$, where
$\alpha$ is the scar amplitude exponent.  By comparison,
the random wave model predicts correlations with a length-scale
of the order of the de Broglie wavelength (i.e.~of the order of
$\hbar$). We illustrate the influence of scarring on the quantum
autocorrelation function with numerical computations for a family
of nonlinearly perturbed cat maps (Bas\'ilio de Matos \& Ozorio de
Almeida 1995; Boasman \& Keating 1995).

The bifurcation of periodic orbits is a characteristic phenomenon
in systems whose phase space is mixed.  It has been conjectured
by Percival (1973) that the energy eigenstates of mixed systems
separate semiclassically into those associated with stable
islands (regular) and those associated with the chaotic sea
(irregular).  The random wave model has recently been extended to
irregular states in mixed systems by B\"acker \& Schubert (2002) 
(BS1 in the references).  The autocorrelation of these states was 
investigated in this context in B\"acker \& Schubert (2002) (BS2 in the 
references). Our results, concerning the length scale of the correlations, are
expected to apply to the autocorrelation of irregular eigenstates
in the vicinity of superscars.

\section{Autocorrelation Formulae}

Our purpose here is to derive semiclassical autocorrelation
formulae for bifurcating periodic orbits which generalize those
obtained by Berry (Berry 1989) (see also Li \& Rouben 2001) for
periodic orbits far from bifurcation.

The autocorrelation function for the eigenfunction $\psi_{n}({\bf
q})$ of a given quantum Hamiltonian corresponding to an energy level
$E_{n}$ is
\begin{equation}
c_n({\bf L})=\langle\psi_n({\bf q+L})\psi_n^*({\bf q})\rangle,
\label{f0a}
\end{equation}
where $\langle\ldots\rangle$ denotes an average over position
${\bf q}$.  We shall be particularly concerned with the case when
the autocorrelation function is averaged over a group of
eigenstates with energy levels near to $E$, that is with
\begin{equation}
C({\bf L},E,\varepsilon)=\sum_{n} c_n({\bf L}) \delta_{\varepsilon}
\left (E-E_{n} \right),
\label{f0b}
\end{equation}
where $\delta_{\varepsilon}(x)$ is a normalized,
Lorentzian-smoothed $\delta$-function of width $\varepsilon$. The
right-hand side of (\ref{f0b}) thus corresponds to a sum over
eigenstates for which $E_{n}$ lies within a range of size of the
order of $\varepsilon$ centred on $E$. Semiclassically, it is
approximately the average of $c_n({\bf L})$ multiplied by
$\overline{d}(E)$, the mean level density, which, for systems with
two-degrees-of-freedom, is given asymptotically by
\begin{equation}
\overline{d}(E) \sim \frac{V(E)}{(2\pi\hbar)^{2}}
\label{f3}
\end{equation}
as $\hbar \rightarrow 0$, where
\begin{equation}
V(E)=\int \delta (E-H({\bf p},{\bf q})) d^{2}{\bf q} d^{2}{\bf p}
\label{f4}
\end{equation}
and $H({\bf p},{\bf q})$ is the classical Hamiltonian.

Our approach follows the one developed by Keating \& Prado
(2001) for the special case ${\bf L}={\bf 0}$. It is based on the
identity
\begin{equation}
G({\bf q+L},{\bf q};E)=\sum_{n}\frac{\psi_{n}({\bf
q+L})\psi^{\ast}_{n}({\bf q})}{E-E_{n}},
\label{f1}
\end{equation}
where $G$ is the Green function.  This implies that
\begin{multline}
\sum_{n} \psi_{n}({\bf q+L})\psi^{\ast}_{n}({\bf q})
\delta_{\varepsilon} \left (E-E_{n} \right
)=-\frac{1}{2\pi i}\left \{G({\bf q+L},{\bf q};E+i\varepsilon)\right. \\ 
\left. - \left(G({\bf q},{\bf q+L};E+i\varepsilon)\right)^{\ast}\right \}.
\label{fg}
\end{multline}
For simplicity, we shall concentrate here on systems with time-reversal 
symmetry.  In this case (\ref{fg}) reduces to
\begin{equation}
\sum_{n} \psi_{n}({\bf q+L})\psi^{\ast}_{n}({\bf q})
\delta_{\varepsilon} \left (E-E_{n} \right
)=-\frac{1}{\pi}\impart G({\bf q+L},{\bf q};E+i\varepsilon).
\label{f2}
\end{equation}
(As will become apparent, our analysis generalizes straightforwardly to 
non-time-reversal-symmetric systems.)
The connection with classical mechanics is achieved using the
semiclassical approximation to the Green function. For systems
with two-degrees-of-freedom, this is
\begin{equation}
G({\bf q+L},{\bf q};E)\approx \frac{1}{i\hbar \sqrt{2\pi i \hbar}}
\sum_{\gamma}\sqrt{\left | D_{\gamma} \right |} \exp{ \left
\{\frac{i}{\hbar} S_{\gamma}({\bf q+L},{\bf
q};E)-\frac{i\pi}{2}\mu_{\gamma} \right \}},
\label{f5}
\end{equation}
where the sum includes all classical trajectories from ${\bf q}$
to ${\bf q+L}$ at energy $E$, $S_{\gamma}$ is the action along
the trajectory labelled $\gamma$,
\begin{equation}
D_{\gamma}=\det{ \left (
\begin{array}{ll}
\frac{\partial^{2}S_{\gamma}}{\partial {\bf q'}\partial {\bf q}} &
\frac{\partial^{2}S_{\gamma}}{\partial {\bf q'}\partial E} \\
\frac{\partial^{2}S_{\gamma}}{\partial E \partial {\bf q}} &
\frac{\partial^{2}S_{\gamma}}{\partial E^{2}}
\end{array} \right )}
\label{f6}
\end{equation}
with ${\bf q'}={\bf q+L}$, and $\mu_{\gamma}$ is the Maslov index
(Gutzwiller 1990).

The autocorrelation formula we seek follows from averaging
(\ref{f2}) with respect to ${\bf q}$. On the left-hand side this
gives $C({\bf L},E,\varepsilon)$.  On the right-hand side, the ${\bf
q}$-average (via the stationary phase condition) semiclassically
selects from the classical orbits with initial and final
positions differing by ${\bf L}$ - henceforth we call such
trajectories ${\bf L}$-trajectories - those for which the initial
and final momenta are equal. Our main concern will be with cases
when $|{\bf L}|$ vanishes in the semiclassical limit.   Then the
classical orbits contributing to the ${\bf q}$-average are close
to periodic orbits.  To first approximation one can thus find
them by linearizing about the periodic orbits. Essentially, when
a periodic orbit is isolated, this corresponds to expanding the
action up to terms that are quadratic in the distance from it.
Specifically, if ${\bf M}_p$ is the monodromy matrix associated
with a periodic orbit $p$, then in a section transverse to the orbit,
in which the coordinate is $y$ and the conjugate momentum is
$p_y$ (we assume here two degrees of freedom, for simplicity),
\begin{equation}
\begin{pmatrix}
  y+L_y \\
  p_y
\end{pmatrix}={\bf M}_p\begin{pmatrix}
  y \\
  p_y
\end{pmatrix},
\label{f6a}
\end{equation}
where $L_y$ is the component of ${\bf L}$ in the direction of the $y$-axis.
For a given choice of $L_y$, this allows one to determine the
coordinates of the associated orbit in the vicinity of $p$ which
contributes to $C({\bf L},E,\varepsilon)$.  Thus one may view the
autocorrelation function semiclassically as a sum of contributions
from the periodic orbits.  The contribution from each periodic
orbit is, within the linear approximation, stationary in $L_y$
when $p_y=0$ (i.e.~on the periodic orbit itself), that is when
$L_y=0$. This may be seen explicitly by expanding the action in
(\ref{f2}) up to quadratic terms around the orbit $p$.  Let $z$
be the coordinate along $p$.  Here and in the subsequent analysis
we assume that periodic orbit contributions to the
autocorrelation function are $z$-independent (this is correct in
determining the $\hbar$ dependence of these contributions when
$|{\bf L}|\rightarrow 0$ as $\hbar \rightarrow 0$, because the
$z$-dependence comes in only via the classical dynamics, except
at self-focal points, which do not affect the leading order
semiclassical asymptotics of the ${\bf q}$-average). The ${\bf
q}$-average may then be computed (it is a Gaussian integral),
giving the contribution from $p$ to be proportional to
$\exp(i\eta_pL^{2}_{y}/\hbar)$, where $\eta_p$ is determined by ${\bf M}_p$.
It is clear that this is
stationary (i.e.~centred) at $L_y=0$. More importantly for us, it
also shows that the contribution from an isolated periodic orbit
has a length-scale that is of the order of $\hbar^{\frac{1}{2}}$
(in contrast to the random wave model component, for which the
length-scale is of the order of $\hbar$). This picture coincides
precisely with Berry's (Berry 1989) for Wigner functions.  The
question we now address is what is the length-scale of the
contribution to the autocorrelation function from bifurcating
(i.e.~non-isolated) periodic orbits.

One might imagine that the autocorrelation formulae for
bifurcating orbits could be obtained easily by expanding the
action in (\ref{f5}) to higher order than quadratic. This is true
for some of the simpler bifurcations (e.g.~the codimension-one
bifurcations of orbits with repetition numbers $r=1$ and $r=2$ -
see the examples in Section 3), but for more
complicated bifurcations it is incorrect, because for these the
linearized map ${\bf M}^{r}_{p}$ is equal to the identity, which
cannot be generated by the action $S({\bf q'},{\bf q};E)$. Thus
it is difficult to build into the semiclassical expression for the
${\bf qq'}$-representation of the Green function a well-behaved
description of the nonlinear dynamics which the linearized map
approximates. The solution to this problem, found by Ozorio de
Almeida \& Hannay (1987), is to transform the Green function to a
mixed position-momentum representation, and this is the approach
we now take.

The method just outlined involves, first, Fourier transforming
$G({\bf q'},{\bf q};E)$ with respect to ${\bf q'}$. This gives
the Green function in the ${\bf q}$ ${\bf p'}$-representation,
$\tilde{G}({\bf p'},{\bf q};E)$ (${\bf p'}$ is the momentum
conjugate to ${\bf q'}$). The semiclassical approximation to
$\tilde{G}$ takes the same form as (\ref{f5}), except that
$S({\bf q'},{\bf q};E)$ is replaced by the ${\bf q}$ ${\bf
p'}$-generating function $\tilde{S}({\bf p'},{\bf q};E)$. $G({\bf
q'},{\bf q};E)$ may then be rewritten, semiclassically, as the
Fourier transform of this expression with respect to ${\bf p'}$.
The result, for the semiclassical contribution to $G({\bf
q+L},{\bf q};E)$ from classical orbits in the neighbourhood of a
bifurcating periodic orbit, takes the following form.

Consider the general case of a codimension-$K$ bifurcation of a
periodic orbit with repetition number $r$. As before, let $z$ be a
coordinate along the orbit at bifurcation, let $y$ be a coordinate
transverse to it, and let $p_{y}$ be the momentum conjugate to
$y$, so that $y$ and $p_{y}$ are local surface of section
coordinates.  We again assume the semiclassical $\hbar$-scaling of the 
length-scale of the contribution from the bifurcating 
orbit to the autocorrelation function to be
$z$-independent.  Let $\Phi_{r,K}(y, p_{y},{\bf x})$ be the
normal form which corresponds to the local (reduced) generating
function in the neighbourhood of the bifurcation (Arnold 1978;
Ozorio de Almeida 1988), where ${\bf x}=(x_{1},x_{2},\cdots,
x_{K})$ are parameters controlling the unfolding of the
bifurcation. Then, up to irrelevant factors, the contribution to
$G({\bf q+L},{\bf q};E)$ is
\begin{equation}
G_{r,K}(y,L_y;{\bf x})\propto\frac{1}{\hbar^2}\int \exp{\left
[\frac{i}{\hbar} \Phi_{r,K}(y,p_{y},{\bf
x})+\frac{i}{\hbar}p_{y}L_y\right ]} dp_{y} \label{f12}
\end{equation}
(as already stated, we are here interested in determining the
$\hbar$-dependence of the length-scale of contributions to the
autocorrelation function, and so have neglected terms in
(\ref{f12}), such as an $\hbar$-independent factor in the
integrand, which do not influence this).  Note that it follows from the
definition of the generating function that the stationary phase
condition with respect to $p_y$ selects classical orbits for
which the initial and final $y$-coordinates differ by $L_y$; that
then the stationary phase condition with respect to $y$ (which
gives the semiclassical contribution to the $y$-average needed to
compute the autocorrelation formula) further selects from these
orbits those whose initial and final momenta, $p_y$, are
identical; and that the orbits contributing at the point where
$G_{r,K}$ is stationary with respect to $L_y$ are those for which
$p_y=0$.

We now follow the steps taken in Keating \& Prado (2001) for the
case when $L_y=0$: first, rescale $y$ and $p_y$ to remove the
$1/\hbar$ factor from the dominant term (germ) of $\Phi_{r,K}$ in
the exponent, and then apply a compensating rescaling of the
parameters $x_{1},x_{2},\cdots, x_{K}$ to remove the
$\hbar$-dependence from the other terms which do not vanish as
$\hbar \rightarrow 0$. Finally, we then rescale $L_y$ to remove
the $\hbar$-dependence of the second term in the exponent in
(\ref{f12}).  Crucially, this term only involves $p_y$, and so if
in the initial rescalings
\begin{equation}
p_y\rightarrow \hbar^{\alpha_{r,K}}p_y
\label{f13}
\end{equation}
then the length scale of the corresponding contribution to the
autocorrelation function is of the order of
$\hbar^{1-\alpha_{r,K}}$. This is our main general result.  It is
important to note that $\alpha_{r,K}$ is precisely the exponent
found for the scarring amplitude in Keating \& Prado (2001).  This
then implies a sum rule for the scaling exponents of scar
amplitudes and the length-scale exponents of wave function
correlations reminiscent of those found in the theory of critical
phenomena.  It is worth emphasizing that the correlation
length-scale is related to the amplitude exponent, rather than
the width exponent (denoted $\omega_{r,K}$ in Keating \& Prado
2001).

Values of $\alpha_{r,K}$ were computed by Keating \& Prado (2001)
for the generic bifurcations found in two-degree-of-freedom
systems using the appropriate normal forms.  For convenience, we
list here the results.  First, for the $r=1$ (i.e.~saddle-node) 
bifurcations which
correspond to cuspoid (i.e.~corank 1) catastrophes,
$\alpha_{1,K}=\frac{1}{2}$.  Values of $\alpha$ for the generic
bifurcations with $r>1$ and $K=1$ are listed in Table 1, and
those for $r>1$ and $K=2$ in Table 2.  Finally,
$\alpha_{r,K}=\frac{1}{2(K+1)}$ in the general case of bifurcations of
orbits for which $r\geq 2K+2$.

\begin{table}[htb]
\begin{center}
\begin{tabular}{|c|c|c|c|} \hline\hline
$r$      & $2$ & $3$ & $\geq 4$ \\
\hline
$\alpha_{r,1}$      & $1/2$     &  $ 1/3$        &     $ 1/4$     \\
\hline\hline
\end{tabular}
\end{center}
\caption{Values of $\alpha_{r,1}$ for the generic, codimension-1
bifurcations.}
\label{table1}
\end{table}

\begin{table}[htb]
\begin{center}
\begin{tabular}{|c|c|c|c|c|c|} \hline\hline
$r$      &  $2$ &   $3$ & $4$  &   $5$ &   $\geq 6$ \\

\hline
$\alpha_{r,2}$      & $1/2$     &   $1/4$        &      $1/4$  &   $1/5$   &   $1/6$       \\
\hline\hline
\end{tabular}
\end{center}
\caption{Values of $\alpha_{r,2}$ for the generic, codimension-2
bifurcations.}
\label{table2}
\end{table}

Note that in all cases listed above $0<\alpha\leq 1/2$, and that
in most cases $\alpha<1/2$.  Thus the length scale of the fringes
$\hbar^{1-\alpha_{r,K}}$ 
is always semiclassically equal to or smaller than for
non-bifurcating orbits, and in most cases is smaller.  In all
cases it is semiclassically larger than the de Broglie wavelength.

The rescaling of $y$ in (\ref{f12}) was defined in Keating \& Prado
(2001) by
\begin{equation}
y\rightarrow \hbar^{\omega_{r,K}}y,
\label{f14}
\end{equation}
where $\omega_{r,K}$ is the scar width exponent.  In the case of
the autocorrelation function, one must average over $y$.  This
leads to the amplitude of $G_{r,K}$ being proportional to
$\hbar^{\alpha_{r,K}+\omega_{r,K}-2}$.  We remark that the
exponent of $\hbar$ in the amplitude is related to amplitude
exponent $\beta$ (Berry {\em et al.} 2000) of the fluctuations in
the spectral counting function associated with the bifurcation in
question, because
\begin{equation}
\alpha_{r,K}+\omega_{r,K}-2=-(\beta_{r,K}+1)
\label{f15}
\end{equation}
(see equation (2.25) in Keating \& Prado 2001).

The rescaling of $x_n$ in (\ref{f12}) was defined  in Keating \& Prado (2001)
 by
\begin{equation}
x_n\rightarrow \hbar^{\sigma_{n,r,K}}x_n.
\label{f16}
\end{equation}
Here, the exponents $\sigma$ describe the range of influence of
the bifurcation in the different unfolding directions $x_{n}$, and
their sum
\begin{equation}
\gamma_{r,K}=\sum_{n=1}^{K}\sigma_{n,r,K}
\label{f17}
\end{equation}
describes the $\hbar$-scaling of the $K$-dimensional ${\bf
x}$-space hypervolume affected by the bifurcation.

\section{Perturbed cat maps}

We now illustrate some of the  ideas described in the previous
section by focusing on a particular example: a family of
perturbed cat maps.  The influence of bifurcations on the scarring of 
eigenfunctions was explored numerically for these systems in Keating \& Prado
(2001).  (The influence of tangent bifurcations on localization in 
wavefunctions was also investigated 
numerically for a different family of maps by Varga {\it et al.} 1999).  We focus here on 
eigenfunction autocorrelations.

The maps we consider are of the form
\begin{equation}
\left ( \begin{array}{l}
         q_{n+1} \\
         p_{n+1}
        \end{array} \right )=
\left ( \begin{array}{ll}
         2 & 1 \\
         3 & 2
        \end{array} \right ) \left ( \begin{array}{l}
                                      q_n \\
                                      p_{n}
                                     \end{array} \right )+
\frac{\kappa S^{'}_{p}(q_n)}{4\pi^2} \left ( \begin{array}{l}
                                  1 \\
                                  2
                                 \end{array} \right ) \mbox{mod 1},
\label{f40}
\end{equation}
where $S^{'}_{p}(q)$ is the first derivative of a periodic
function with period $1$, $\kappa$ is a parameter determining the
size of the perturbation, and $q$ and $p$ are coordinates on the
unit two-torus which are taken to be a position and its conjugate
momentum. We will consider two cases: $S_p(q)=\sin{(2\pi q)}$ and
$S_p(q)=\cos{(2\pi q)}$.

These maps are Anosov systems.  In both cases, for 
$\kappa \leq \kappa_{\rm
max}=(\sqrt{3}-1)/\sqrt{5}\approx 0.33$ they are completely
hyperbolic and their orbits are conjugate to those of the map
with $\kappa=0$ (i.e.~there are no bifurcations). Outside this
range, bifurcations occur, stable islands are created, and the
dynamics becomes mixed (Berry {\em et al.} 1998, Keating \& Prado
2001).

The quantization of maps like (\ref{f40}) was developed by
Hannay \& Berry (1980), when $\kappa=0$, and Bas\'ilio de
Matos \& Ozorio de Almeida (1995) for non-zero $\kappa$.
The quantum kinematics associated with a phase space that
has the topology of a two-torus restricts Planck's constant
to taking inverse integer values.  The integer in question,
$N$, is the dimension of the Hilbert space of admissible
wavefunctions.  With doubly  periodic boundary conditions
(see, for example, Keating {\em et al.} 1999), these
wavefunctions in their position representation have support
at points $q=Q/N$, where $Q$ takes integer values between 1
and $N$.  They may thus be represented by $N$-vectors with
complex components.  The quantum dynamics is then generated
by an $N \times N$ unitary matrix ${\bf U}$ whose action on
the wavefunctions reduces to (\ref{f40}) in the classical
limit; for example
\begin{equation}
U_{Q_2,Q_1}=\frac{1}{\sqrt{iN}}\exp \left [ \frac{2\pi i}
{N} (Q_1^2-Q_1Q_2+Q_2^2)+\frac{iN}{2\pi}\kappa\sin (2\pi Q_1/N+\pi\nu/2) \right ].
\label{f41}
\end{equation}
where $\nu=0$ gives the sine perturbation and $\nu=1$ the cosine 
perturbation. This matrix plays
the role of the Green function of the  time-dependent
Schr\"odinger equation for flows.

Denoting the eigenvalues of ${\bf U}$ by
${\rm e}^{i\theta_n}$ and the corresponding eigenfunctions by
$\Psi_n(Q)$, we have, using the fact that the map (\ref{f40}) is invariant 
under time-reversal and hence that ${\bf U}$ is symmetric, that
\begin{multline}
C(L, \theta, \epsilon)=\sum^{N}_{n=1} c_{n}(L)
\delta_{\varepsilon}(\theta-\theta_{n})
=\delta_{L,0}+\frac{1}{N}\rpart\sum^{N}_{Q=1}
\sum^{\infty}_{k=1}U_{Q+L,Q}^{k}\exp{(-i\theta k -\varepsilon k)}
\label{f42}
\end{multline}
where
\begin{equation}
c_{n}(L)=\frac{1}{N}\sum_{Q=1}^{N}\Psi_{n}(Q+L)\Psi^{\ast}_{n}(Q)
\label{f43a}
\end{equation}
is the eigenvector autocorrelation function and
\begin{equation}
\delta_{\varepsilon}(x)=\frac{1-e^{-\varepsilon}\cos{x}}{1+e^{-2\varepsilon}-2e^{-\varepsilon}\cos{x}}
\label{f43}
\end{equation}
is a periodized, Lorentzian-smoothed $\delta$-function of width
$\varepsilon$ (Keating 1991).  Equation (\ref{f42}) is the
analogue for quantum maps of (\ref{f2}).  $C(L, \theta, \epsilon)$
corresponds, approximately, to $N$ times the local $n$-average
(over a range of size of the order of $\varepsilon$) of
$c_{n}(L)$.  For $\varepsilon$ large enough, the dominant
contribution to (\ref{f42}) comes from the $k=1$ term in the sum
on the right. We thus define
\begin{equation}
c(L, \theta, \epsilon)=\frac{1}{N}\sum^{N}_{Q=1}
U_{Q+L,Q}\exp{(-i\theta -\varepsilon )}.
\label{f44a}
\end{equation}
and, for computational purposes,
\begin{equation}
c_r(L, \theta, \epsilon)=\frac{1}{N}\rpart  \sum^{N}_{Q=1}
U_{Q+L,Q}\exp{(-i\theta -\varepsilon )}.
\label{f44are}
\end{equation}

We may substitute (\ref{f41}) directly into (\ref{f44a}).  In the
semiclassical limit, as $N \rightarrow \infty$, the $Q$-average
selects regions close to stationary points of the phase of
(\ref{f41}).  When $L=0$ these
stationary points coincide with the positions of the fixed points
of the classical map (\ref{f40}); otherwise the stationary
solutions correspond to $l$-trajectories - in this case classical
trajectories with initial and final coordinates separated by
$l=L/N$ and with initial and final momenta equal to each other.
These satisfy
\begin{equation}
q_j(l)=\frac{1}{2} \left ( j-l-\frac{\kappa}{2\pi}\cos{ (2\pi
q_j+\nu\frac{\pi}{2}}) \right)
\label{f45a}
\end{equation}
for integers $j$ such that $0 \leq q_j(l) < 1$ (see, for example,
Boasman \& Keating 1995).
When $l\rightarrow 0$ (which is the most important regime, 
given that we wish to focus on correlation length-scales that vanish semiclassically) the
$l$-trajectories obviously lie close to fixed points of the map.  

Expanding the phase of (\ref{f41})
around ${q}_j$ up to cubic terms gives
\begin{multline}
U_{Q+L,Q}\approx \frac{1}{\sqrt{iN}}\exp \left [ 2\pi i N S_j+ \pi iN \left( 2-\kappa \sin \left( 2\pi q_j+ 
\nu\frac{\pi}{2} \right)\right) y^2 \right. \\ 
\left.  -\frac{2\pi^2 iN}{3}\kappa 
\cos \left( 2\pi q_j+\nu\frac{\pi}{2} \right) y^3
\right ], 
\label{f45}
\end{multline}
where
\begin{equation}
y=q-q_j(l)
\label{f46}
\end{equation}
and $2\pi N S_j$ denotes the phase evaluated at $q_j(l)$.
Provided that $2-\kappa \sin (2\pi  q_j+\nu\pi/2) \ne 0$, this
approximation is dominated by the quadratic term in the exponent
when $y$ is small.  It thus describes complex-Gaussian fringes
around the classical $l$-trajectories with a length-scale (in
terms of $y$) of the order of $N^{-1/2}$. 

As discussed in Section 2, 
the centres of the
peaks of the autocorrelation function are given by contributing
$l$-trajectories that start at $q$, end at $q+l$, and satisfy $p'=p=0$ in the 
first iterate of the map (\ref{f40}). From (\ref{f40}), the peak 
centres are thus 
solutions of
\begin{equation}
l=q+\frac{\kappa}{2\pi}\cos{(2\pi  q +\nu\pi/2)}-n
\label{f46a}
\end{equation}
and
\begin{equation}
3q+\frac{\kappa}{\pi}\cos{(2\pi  q+\nu\pi/2)}-m=0
\label{f46aa}
\end{equation}
where  $m$ and $n$ are integers. For the linear mapping, when
$\kappa=0$, the peaks are centred at $l=q=m/3$ with  $m=0,1,2$. 
These peaks may be seen in Figure 1, where the dashed
line represents an evaluation of (\ref{f44are}) and the bold line a
convolution of the data with a normalized Gaussian of width 0.007.

For $\kappa>0$ the peak location equations, (\ref{f46a}) and (\ref{f46aa}), 
can be solved numerically. As $\kappa$
increases bifurcations of periodic orbits take place. It is important 
at this stage to distinguish
between the birth of new periodic orbits as the perturbation parameter $\kappa$ is
varied  (the usual bifurcation) and the birth of new
$l$-trajectories as $l$ varies (a process which we call an
$l$-caustic).  As $l \rightarrow 0$, the $l$-trajectories
are close to periodic orbits.  $l$-caustics coincide with periodic orbit 
bifurcations in the limit.

Our main purpose now is to illustrate quantitatively the influence of 
bifurcating periodic orbits  on the autocorrelation function. 
We shall study in detail two bifurcations:
one a first order (tangent) bifurcation, and the other 
a second order bifurcation. In order to do so, 
we will treat separately the sine and cosine
perturbations.

\subsection{Sine perturbation $\nu=0$}

For $\kappa>0$, the positions of the peaks in the autocorrelation function 
deviate from those ($0$, $1/3$ and $2/3$) of the unperturbed map
($\kappa=0$). The real part of the autocorrelation  function for  $\kappa=1$ is shown in Figure 2 (a).  The
peak close to $l=0$ comes from the contribution of 
the period-1 fixed point $j=0$ in (\ref{f45a}).  Using a
stationary phase approximation with respect to $L$ in (\ref{f45})
gives the position of the peak to be
\begin{equation}
l\approx \frac{\kappa}{\pi}
\frac{\cos{(2\pi q_0(0))}}{
 3-\kappa \sin(2\pi q_0(0))}    {\mbox{ mod 1}}
\label{f46b}
\end{equation}
in the limit as $l \rightarrow 0$.
For  $\kappa \sin (2\pi  q_j)) < 2$, the second order
term  in the Taylor expansion (\ref{f45}) dominates  
and the semiclassical
formula for the autocorrelation function is simply
\begin{equation}
c_r(L,\theta ,\epsilon)\approx\sum^{1}_{j=0}\exp{(-\varepsilon)}\frac{\cos{(2\pi N S_{j}-\theta
+(\zeta-1)\pi/4 )}}{\sqrt{|2-\kappa \sin{( 2\pi  q_{j}(l)} )|}}
\label{f46c}
\end{equation}
where
$\zeta= {\mbox{sign}}(2-\kappa\sin{(2\pi q_j(l))})$ and
\begin{equation}
S_{j}=\frac{3l^2}{4}+j\frac{l}{2}-j\frac{1}{4}
+\frac{\kappa}{4\pi^2}\sin{(2\pi  q_j(l))}
+\frac{\kappa^2}{16\pi^2}\cos^2{(2\pi  q_j(l) )}.
\label{f46d1}
\end{equation}
In figure 2 (b) we plot the difference between the exact expression 
(\ref{f44are}) 
and the semiclassical approximation (\ref{f46c}).

For larger  $\kappa$, keeping terms only up to second order in  the  approximation
(\ref{f45}) fails in two ways.
First,  $2-\kappa \sin (2\pi  q_j(l))$  vanishes for some values of
$l$ and higher terms of the expansion of the action need to be taken
into account. It is clear in Figure 3, where the predicted peak-centres (obtained by
solving (\ref{f46a}) and (\ref{f46aa})) are plotted,  that for $\kappa$ around $3.0$  new
$l$-trajectories contribute to the autocorrelation  with  a peak
around $l=0.5$. Although these $l$-caustics are an  additional
complication, they are in fact unimportant if the object is to compute the
contribution of bifurcating periodic orbits which takes place for
$l$ around zero (recall that this is the region that is semiclassically 
interesting for correlation 
length-scales that vanish with $\hbar$).

Second, when  $l \rightarrow 0$,
$2-\kappa \sin (2\pi  q_j(0)) = 0$ at periodic orbit bifurcations. 
Then again the quadratic term in (\ref{f45})
vanishes, and the fringe structure comes from the cubic term.
It thus has a $y$-length-scale of the order of $N^{-1/3}$.  In the language
of Section 2, this corresponds to a codimension-one bifurcation of a
periodic orbit with $r=1$ (a tangent bifurcation).
The   $j=0$ and $j=1$ terms each correspond to a
single unstable fixed point if the condition $2-\kappa \sin (2\pi  q_j(l)) \ne 0$
 is satisfied.  Note that at both $l$-caustics and tangent bifucations, the semiclassical 
asymptotics is determined by the cubic term in the
expansion of the action; hence both contributions are of the same order in $\hbar$.

Figure 4 provides a panorama of the situation.
Here, the absolute value of the Gaussian smoothed autocorrelation function $c$ is plotted.  
The result should be compared with the predicted peak-centres shown in Figure 3.
The peak coming from the bifurcating periodic orbits is around $\kappa=6$
and $l=0$. The other similar peaks are all associated with the
$l$-caustics.  It is clear that their amplitudes are  much
larger than those of the peaks associated with isolated $l$-trajectories.

For the sine perturbation, the first bifurcation occurs when
$\kappa=\kappa^{*}=5.943388$.  At this
parameter value, two new degenerate solutions of (\ref{f45a})
appear  for both $j=0$ and $j=1$, corresponding in each case to the birth of a
pair of fixed points, one stable and
the other unstable. For $l=0$ the contributions of these bifurcating orbits when
$\kappa=\kappa^{\ast}$ are:
  \begin{equation}
c_r(0,\theta ,\epsilon)\approx \sum^{1}_{j=0}\frac{N^{1/6}}{\Gamma(2/3)}\left ( \frac{4\pi}{9\kappa^{\ast}} \right)^{1/3}
\exp{(-\varepsilon)} \frac{\cos{(2\pi i N {\overline{S}}_{j}-\theta -\pi/4)}}{|\cos{(2\pi {\overline{q}}_{j})}|^{1/3}}
\label{f46d}
\end{equation}
where ${\overline{S}}_{j}=(S({q_{+}}_{j})+S({q_{-}}_{j}))/2$ is the average of the
actions of the two bifurcating fixed points at ${q_{+}}_{j}$ and  ${q_{-}}_{j}$.
More generally, the semiclassical contributions of the 
real orbits plus the
complex orbits  to the autocorrelation function
give:
\begin{multline}
c_r(L,\theta ,\epsilon)\approx \sum^{1}_{j=0}\exp{(-\varepsilon)}
\frac{\cos{(2\pi N S_{j}-\theta
+(\sigma-1)\pi/4 )}}{\sqrt{|2-\kappa \sin{( 2\pi  q_j(l)} )|}}+ \\
\rpart{\sum^{1}_{j=0}
\exp{({-\varepsilon})}N^{1/6}\frac{{\mbox{Ai}}\left [-(2\pi N |\alpha|)^{2/3}\right ]}
{2\pi|\alpha|^{1/3}}\sigma^2 \exp{(2\pi i N {\overline{S}}_{j}-i\theta -i\pi/4)}}.
\label{f46e}
\end{multline}
Here $\sigma=(q_{+}-q_{-})/2$ and $|\alpha|=3|\Delta S|/(2\sigma^3)$, where 
$\Delta S=(S(q_{+})-S(q_{-}))/2$. Figure 5 (a) shows the 
autocorrelation function for
$\kappa=\kappa^{\ast}$. In Figure 5 (b) the deviation of the semiclassical approximation
(\ref{f46e}) from the exact expression
(\ref{f44are}) is plotted.

\subsection{Cosine perturbation $\nu=1$}

Setting $\nu=1$ in (\ref{f41}) and the following equations
allows us now to discuss a second order bifurcation.  (Such 
bifurcations occur for the second iterate of the map with the sine 
perturbation, and so quantum mechanically influence $U^{2}(Q,Q)$.  In the
case of the cosine perturbation, they occur for fixed points of the map, 
and so influence $U(Q,Q)$.  This makes them easier to study 
numerically (Mende (1999).)

As the order of the first non-vanishing derivative of the action 
$S_{p}(q)$ determines the
order of the bifurcation, one can see that there are no higher bifurcations
than those of second order for this perturbation. For $j=0$, $q=0$ and
$\kappa=2$, a bifurcation of order two occurs. For $j=1$ there is no such
bifurcation: in this case only tangent bifurcations occur, the first one  
at $\kappa \approx 9.208$ (Mende 1999).
As the tangent bifurcation was discussed previously, we will concentrate
 on the bifurcation of order two.  For this, when $\kappa<2$
there is one real unstable
fixed point (called the central fixed point), while for $\kappa>2$ there are 
two real fixed points (satellites) in addition to the central fixed point, which
is now stable.
The scarring of the wavefunctions caused by this bifurcation is shown as 
a dashed line is Figure 6, where
$\sum_{n}|\Psi_{n}(Q)|^{2}\delta_{\varepsilon}(\theta-\theta_n) -1$
is plotted for (a) $\kappa=1$ (one real unstable fixed point ($j=0$) at 
$q=0$ and another real  unstable fixed point ($j=1$) at  $q=0.5$), 
(b) $\kappa=\kappa^{\ast}=2$ (where the $j=0$ fixed point
bifurcates), and (c) $\kappa=3$  (three real  $j=0$ fixed points - unstable satellites at $q=-0.23$ and $q=0.23$ and the central fixed point at $q=0$, which
is now stable - and the unstable $j=1$ fixed point at $q=0.5$).
A convolution of the data with a normalized Gaussian
of width 0.02 is also shown (bold line).

In Figure 7, $|\sum_{Q=1}^{N}U(Q,Q)|^2=|{\rm Tr}U|^2 $ is plotted as a
function of $\kappa$ and $N$.  One sees the contribution of the
second order bifurcation of the $j=0$ fixed point around $\kappa= 2$
and that of the tangent bifurcation of the $j=1$ fixed point around $\kappa=9$.
The amplitude associated with the second order bifurcation is clearly
semiclassically larger. 
Similarly, the amplitude of the autocorrelation function is larger at
these points (compare (\ref{f47}) below with (\ref{f46d})). The predicted positions of the 
peaks in the autocorrelation function (obtained by solving (\ref{f46a}) and 
({\ref{f46aa}), when $\nu=1$) are
shown in Figure 8.   Here again, for $\kappa=0$ the peaks are centred at
$l=0\equiv 1$, $l=1/3$ and $l=2/3$, as previously described. 
As $\kappa$ increases, 
these peaks deviate from their original positions, but  the
autocorrelation result  for isolated points, the analogue of (\ref{f46c}), is still valid.
For $\kappa$ around 2, where the second order bifurcation occurs,
the approximation (\ref{f45}) breaks down and  an expansion of the phase
of (\ref{f41}) around ${q}_j$  up to  fourth order terms is necessary.
The contribution of the bifurcating fixed point  ($j=0$)  when 
$\kappa=\kappa^{\ast}=2$ and $L=0$ is 
\begin{equation}
c_r(0,\theta ,\epsilon) \approx \frac{N^{1/4}}{2\sqrt{2} \Gamma(3/4)}
\left ( \frac{3\pi}{\kappa^{\ast}} \right)^{1/4}
\exp{(-\varepsilon)} \cos{(2\pi i N {S_0}-\theta -\pi/4)}
\label{f47}
\end{equation}
where ${S_0}$ is the
action of central orbit at $q=0$. More generally, for $\kappa \lesssim 9 $ 
the semiclassical 
approximation to the autocorrelation function is
\begin{multline}
c_r(L,\theta ,\epsilon)\approx \left \{
\frac{\cos{(2\pi N S_{1}-\theta +(\zeta-1)\pi/4 )}}
{\sqrt{|2-\kappa \cos{( 2\pi  q_1(l))} |}} \right. \\
+ \left .\rpart{ \frac{\pi}{2}N^{1/2}q_s(l)
\left [ {\mbox J}_{-1/4}(z)\exp{(i \sigma_1)}+
{\mbox{sign}}(2-\kappa){\mbox J}_{1/4}(z)
\exp{(i\sigma_2)}\right ]} \right \} \exp{(-\varepsilon)}.
\label{f48}
\end{multline}
Here $q_s(l)$ is the distance of the satellite points from 
the origin, $z=\pi N \alpha  q_{s}^{4}/4$,
where $\alpha=4(- q_s^2+\kappa (1-\cos{(2\pi   q_{s})}))/ q_s^4$,
$\sigma_1=-\theta -\pi/8+2\pi N S_0 -\pi N \alpha  q_s^4/4$, and
$\sigma_2=-\theta +5\pi/8+2\pi N S_0 -\pi N \alpha  q_s^4/4$.
 
In Figure 9 (a) the real part of the autocorrelation  function for
$\kappa=\kappa^{\ast}=2$ is plotted and in (b) the deviation of the semiclassical 
approximation (\ref{f48}) from the exact 
expression (\ref{f44are})  is shown. 
In Figure 10 the absolute
value of the  convolution of the  autocorrelation function (\ref{f44a}) with a
normalized Gaussian  is plotted as a function of $\kappa$ and $l$, showing the
consistency of the theory: higher amplitudes near $l$-caustics and
closed agreement with the predicted positions of the peak-centres (Figure 8).  
Finally, in Figure 11 the absolute value of the real part of the
autocorrelation function (\ref{f44are}) 
at $L=0$ (convolved with a Gaussian in $\kappa$ of
width 0.1) is plotted.  From this one sees clearly the influence of the
two bifurcations: the tangent bifurcation at $\kappa \approx 9.208$ and the 
second order bifurcation at $\kappa=2$.

\section{Acknowledgements}

AB thanks the School of Mathematics, University of Bristol, and the 
Basic Research Institute in the Mathematical Sciences (BRIMS), Hewlett-Packard 
Laboratories, Bristol, for hospitality during the period when this work was started.  
SDP is grateful to Martin Sczyrba for helpful discussions, and to the
School of Mathematics at the University of Bristol for hospitality and
 financial support.  Support is also acknowledged 
from the Deutsche Forschungsgemeinschaft (for AB) and  
the Funda\c{c}\~ ao de Amparo
\`a Pesquisa do Estado do Rio Grande do Sul - FAPERGS (for SDP).

\ \


\noindent {\Large{ \bf References}}


\ \

\noindent Arnold, V.I. 1978 {\em Mathematical Methods in Classical
Mechanics.} Springer.

\ \

\noindent B\"acker, A. \& Schubert, R. 2002 Amplitude distribution
of eigenfunctions in mixed systems. {\em J. Phys.} A {\bf 35}, 527-538 (BS1).

\ \

\noindent B\"acker, A. \& Schubert, R. 2002 Autocorrelation function
of eigenstates in chaotic and mixed systems. 
{\em J. Phys.} A {\bf 35}, 539-564 (BS2).

\ \

\noindent Bas\'ilio de Matos, M. \& Ozorio de Almeida, A.M. 1995
Quantization of Anosov maps.  {\em Ann. Phys.} {\bf 237}, 46-65.

\ \

\noindent Berry, M.V. 1977  Regular and irregular semiclassical wavefunctions
   {\em J. Phys.} A {\bf 10}, 2083-2091.

\ \

\noindent Berry, M.V. 1989, Quantum scars of classical closed orbits in phase
space. {\em Proc. R. Soc. Lond.} A  {\bf 243}, 219-231.

\ \

\noindent Berry, M.V,  Keating, J.P \& Prado, S.D. 1998 Orbit bifurcations and
spectral statistics. {\em J. Phys.} A {\bf 31}, L245-254.

\ \

\noindent Berry, M.V., Keating, J.P. \& Schomerus, H. 2000 Universal twinkling
exponents for spectral fluctuations associated with mixed chaology.
{\em Proc. R. Soc. Lond.} A {\bf 456}, 1659-1668.

\ \

\noindent Boasman, P.A. \& Keating, J.P. 1995 Semiclassical asymptotics of
perturbed cat maps.  {\em Proc. R. Soc. Lond.} A {\bf 449}, 629-653.

\ \

\noindent Bogomolny, E.B. 1988 Smoothed wavefunctions of chaotic quantum systems. {\em Physica} D {\bf 31}, 169-189.

\ \

\noindent Gutzwiller, M.C. 1990 {\em Chaos in Classical and Quantum Mechanics} (New York: Springer).

\ \

\noindent Hannay, J.H. \& Berry, M.V. 1980 Quantization of linear maps on
the torus -- Fresnel diffraction by a periodic grating.  {\em Physica} D
{\bf 1}, 267-290.

\ \

\noindent Heller, E.J. 1984 Bound state eigenfunctions of classically
chaotic Hamiltonian systems - scars of periodic orbits.
{\em Phys. Rev. Lett.} {\bf 53}, 1515-1518.

\ \

\noindent Kaplan, L. 1999 Scars in quantum chaotic wavefunctions.
{\em Nonlinearity} {\bf 12}, R1-R40.

\ \

\noindent Keating, J.P. 1991 The cat maps: quantum mechanics and classical
motion.  {\em Nonlinearity} {\bf 4}, 309-341.

\ \

\noindent Keating, J.P., Mezzadri, F. \& Robbins, J.M. 1999 Quantum
boundary conditions for torus maps.  {\em Nonlinearity} {\bf 12}, 579-591.

\ \

\noindent Keating, J.P. \& Prado, S.D. 2001 Orbit bifurcations and the
scarring of wave functions.  {\em Proc. R. Soc. Lond.} A {\bf 457}, 1855-1872.

\ \

\noindent Li, B. \& Rouben, D.C. 2001 Correlations of chaotic eigenfunctions: a semiclassical analysis.  {\em J. Phys.} A {\bf 34}, 7381-7391.

\ \

\noindent Mende, M. 1999 Periodic orbit bifurcations and spectral statistics.  M.Sc. Thesis, University of Bristol.

\ \

\noindent Ozorio de Almeida, A.M. 1988 {\em Hamiltonian Systems: Chaos
and Quantization.} Cambridge University Press.

\ \

\noindent Ozorio de Almeida, A.M. \& Hannay, J.H. 1987 Resonant periodic orbits
and the semiclassical energy spectrum. {\em J. Phys} A {\bf 20},
5873-5883.

\ \

\noindent Percival, I.C. 1973 Regular and irregular spectra.
{\em J. Phys.} B {\bf 6}, L229-L232.

\ \

\noindent Varga, I., Pollner, P. \& Eckhardt, B. 1999 Quantum localization near
bifurcations in classically chaotic systems.  {\em Ann. Phys.} (Leipzig)
{\bf 8}, SI265-SI268.


\newpage

\begin{figure}[htb]
\begin{center}
\leavevmode
\psfig{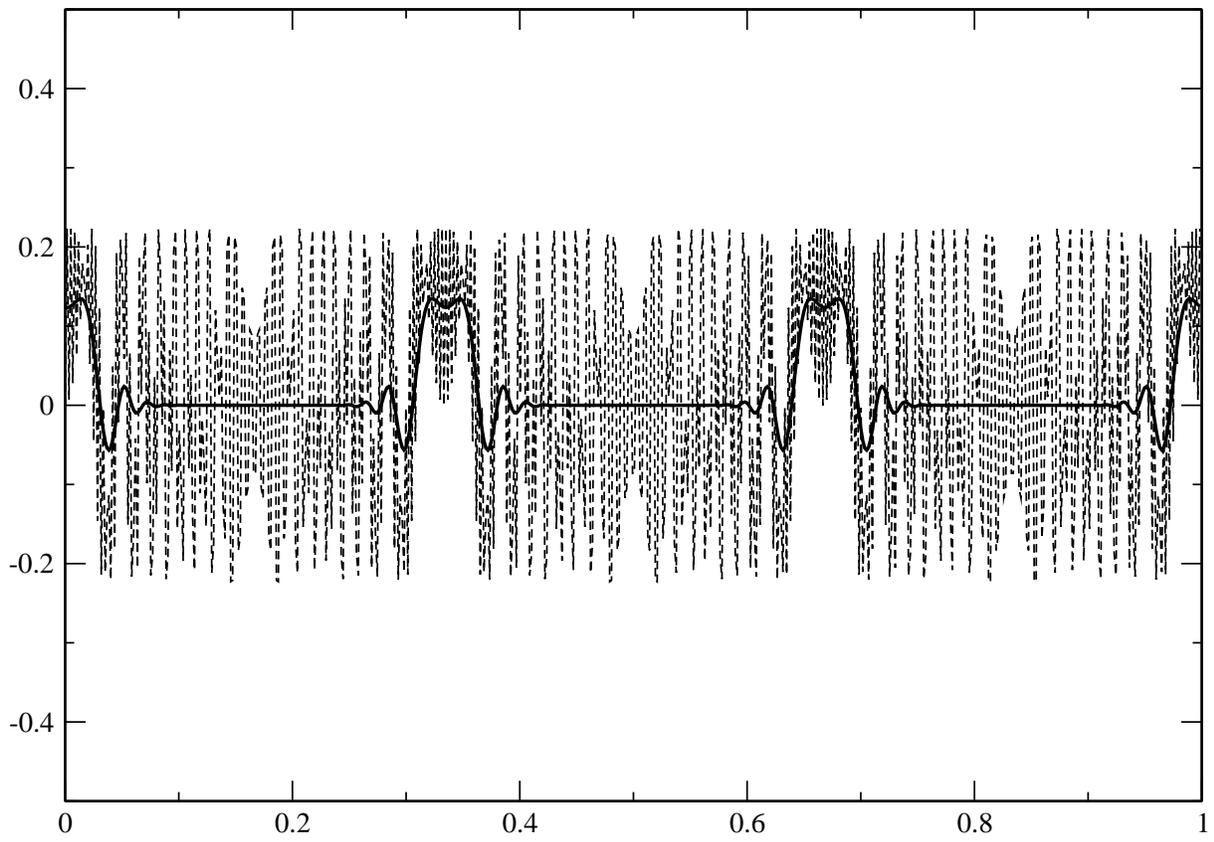}
\end{center}
\caption{$c_r(L,\theta ,\epsilon)$ plotted as a function of $l=L/N$ with 
$N=597$, $\kappa=0.0$ (i.e.~no perturbation),
 $\epsilon=1.5$ and $\theta=\pi/4$ (dashed line). The bold line  
represents a convolution of the data with a normalized Gaussian of
width 0.007.}
\label{fig:f1}
\end{figure}

\begin{figure}[htb]
\begin{center}
\leavevmode
\psfig{figure=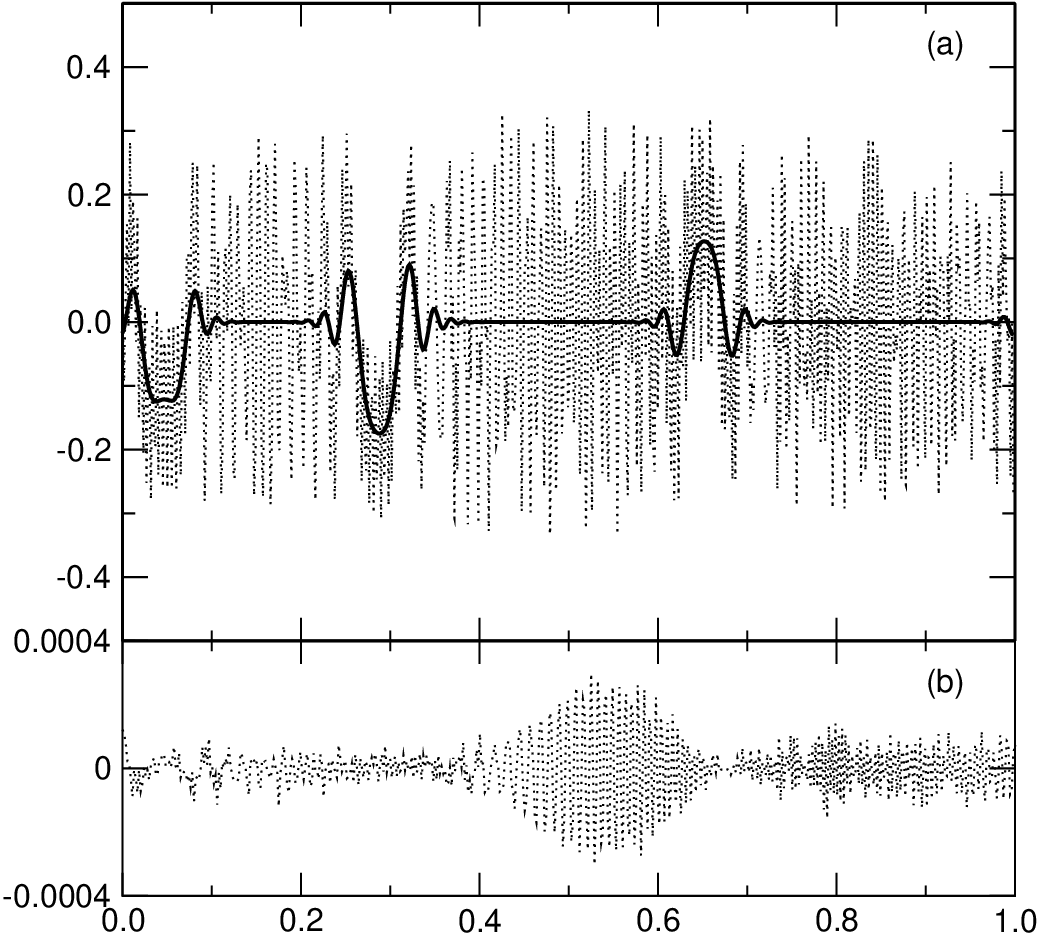,width=16.0cm,angle=0}
\end{center}
\caption{(a) $c_r(L,\theta ,\epsilon)$ plotted as a function of $l=L/N$ 
for the sine perturbation with 
$N=597$, $\kappa=1.0$,
 $\epsilon=1.5$ and $\theta=\pi/4$ (dashed line). The bold line  
represents a convolution of the data with a normalized Gaussian of
width 0.007. In (b) the difference between the exact
expression (\ref{f44are}) and the semiclassical 
approximation (\ref{f46c}) is plotted.}
\label{fig:f2}
\end{figure}

\begin{figure}[htb]
\begin{center}
\leavevmode
\psfig{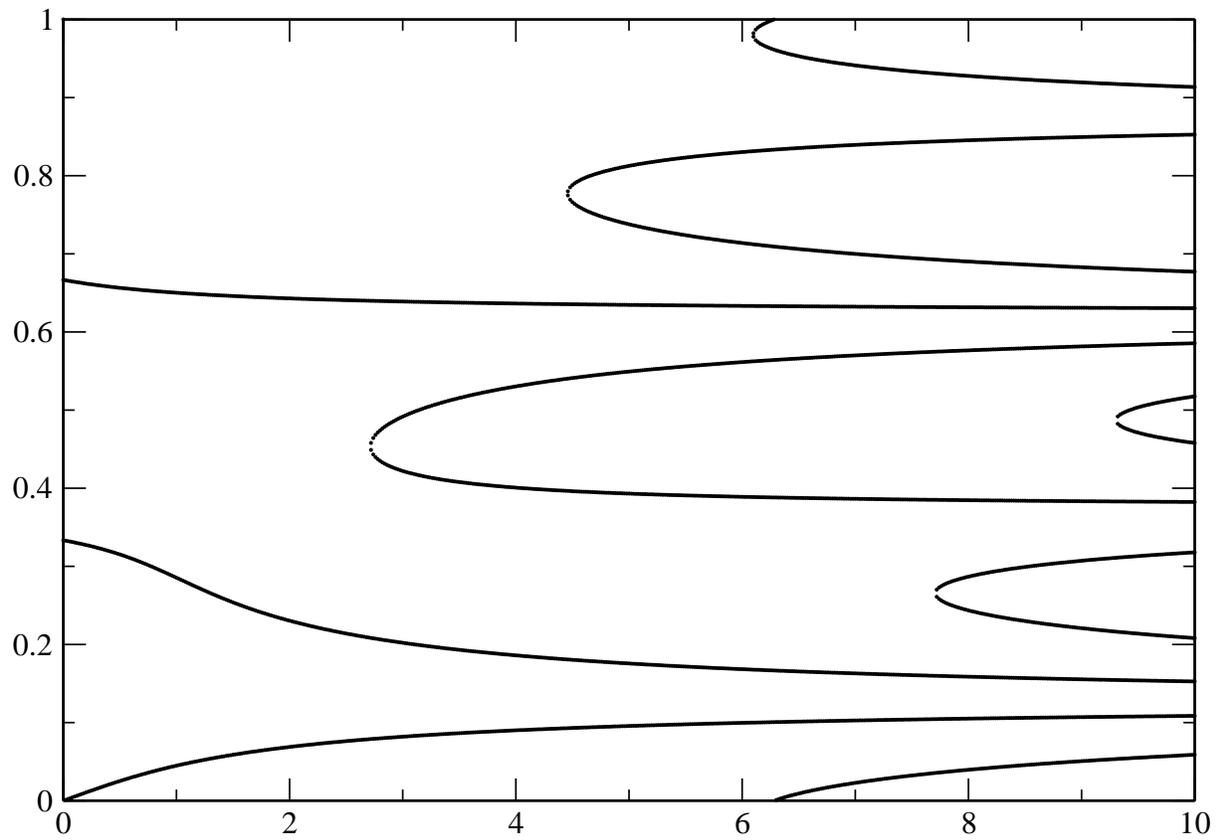}
\end{center}
\caption{Predicted peak-centres in the 
autocorrelation function due to $l$-trajectories, computed by solving 
(\ref{f46a}) and (\ref{f46aa}), for the sine perturbation.}
\label{fig:f3}
\end{figure}

\begin{figure}[htb]
\begin{center}
\leavevmode
\psfig{figure=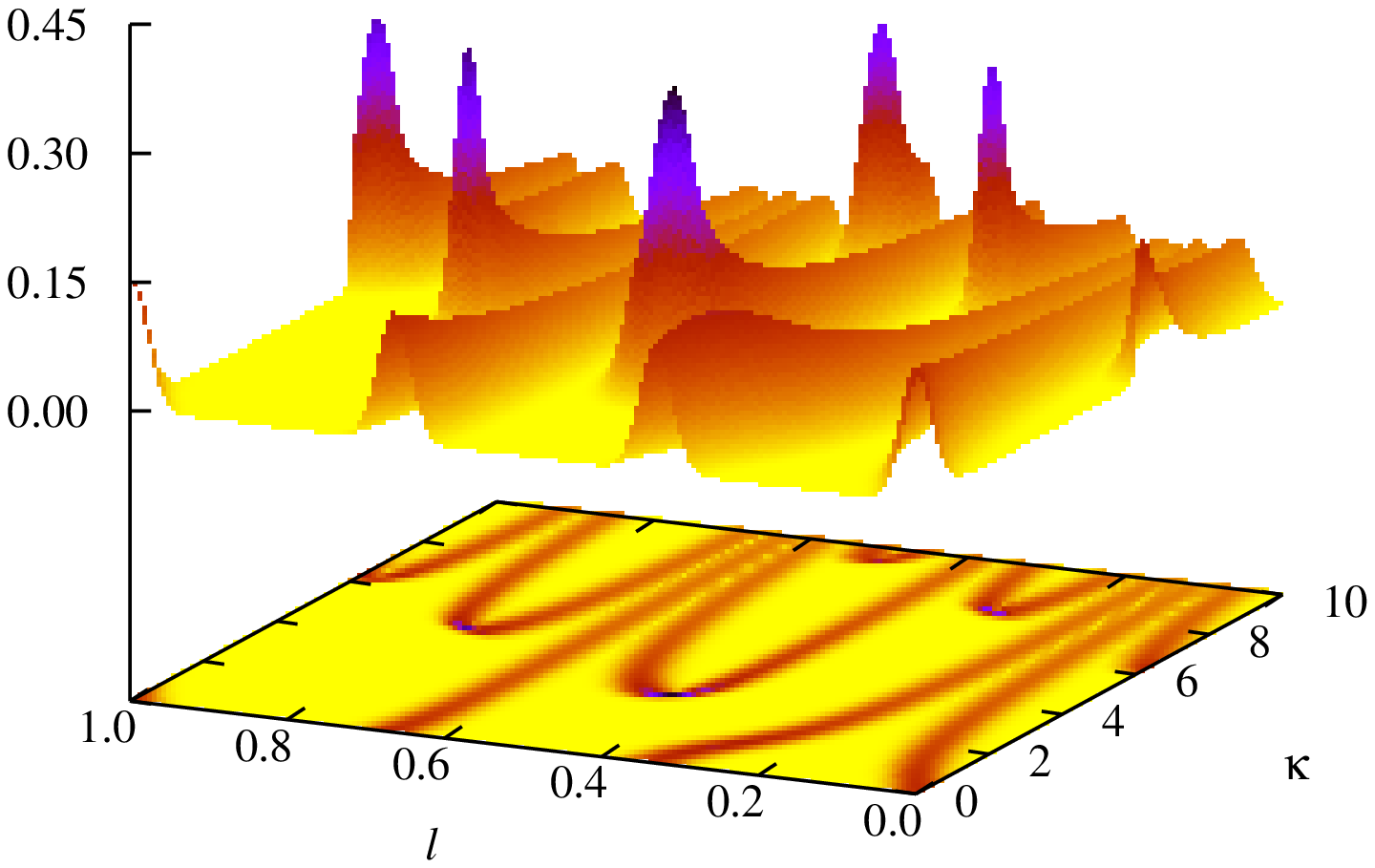,width=16.0cm,angle=0}
\end{center}
\caption{The absolute value of the convolution of the
 autocorrelation function (\ref{f44a}) with a normalized Gaussian of width 0.01,  plotted as a
function of $\kappa$ and $l=L/N$ for the sine perturbation with $N=597$, $\epsilon=1.5$ 
and $\theta=\pi/4$. 
A projection is shown underneath.  Compare the positions of the ridges with
the theoretical curves shown in Figure 3.  Note also the enhanced  
amplitude near the bifurcation and the $l$-caustics.}
\label{fig:f4}
\end{figure}

\begin{figure}[htb]
\begin{center}
\leavevmode
\psfig{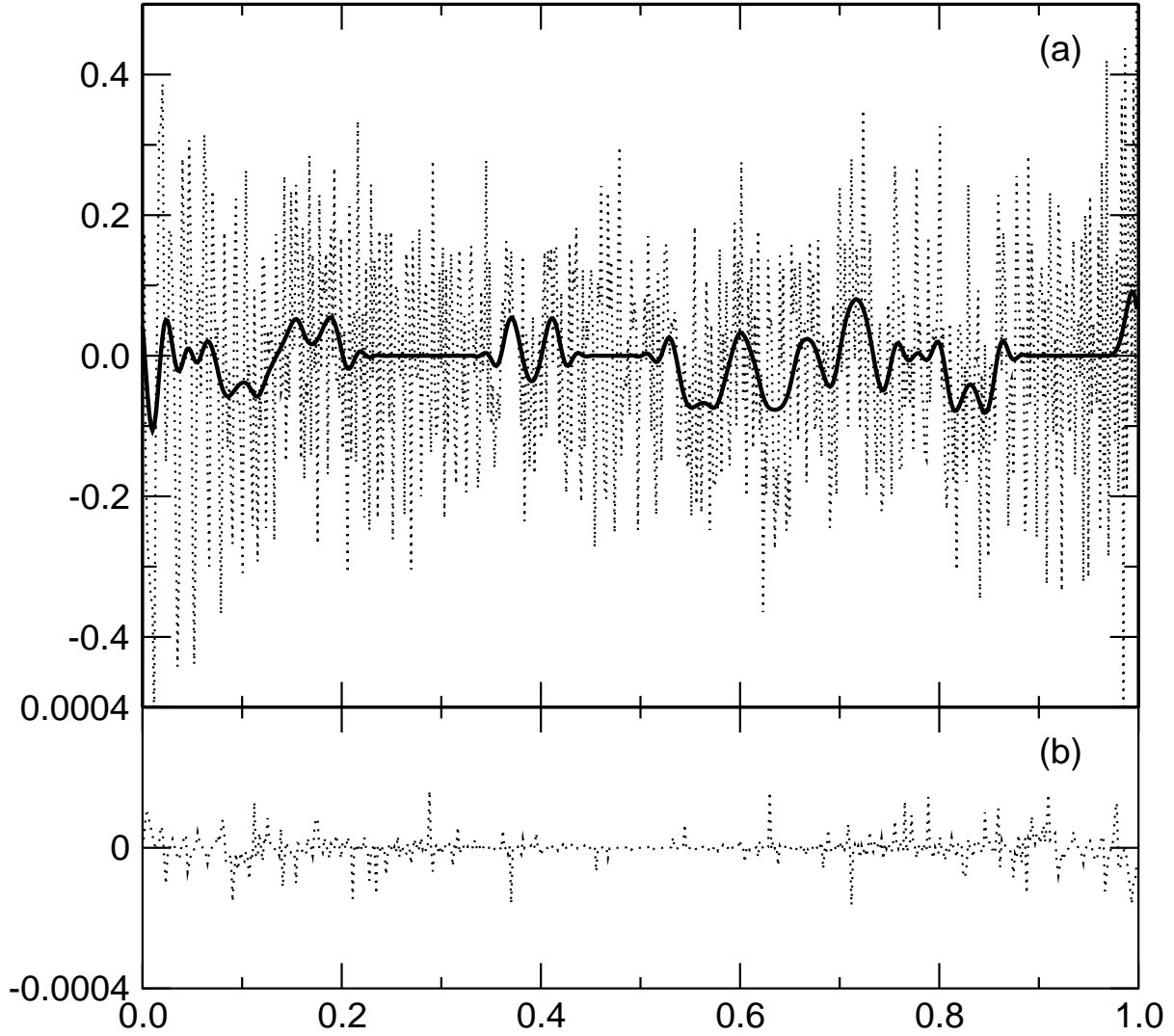}
\end{center}
\caption{(a) $c_r(L,\theta ,\epsilon)$ plotted as a function of $l=L/N$ 
for the sine perturbation with 
$N=597$, $\kappa=5.943388$,
 $\epsilon=1.5$, $\theta=\pi/4$ and $\theta=\pi/4$ (dashed line). 
The bold line  
represents a convolution of the data with a normalized Gaussian of
width 0.007. In (b) the difference between the exact
expression (\ref{f44are}) and the semiclassical approximation 
(\ref{f46e}) is plotted.}
\label{fig:f5}
\end{figure}

\begin{figure}[htb]
\begin{center}
\leavevmode
\psfig{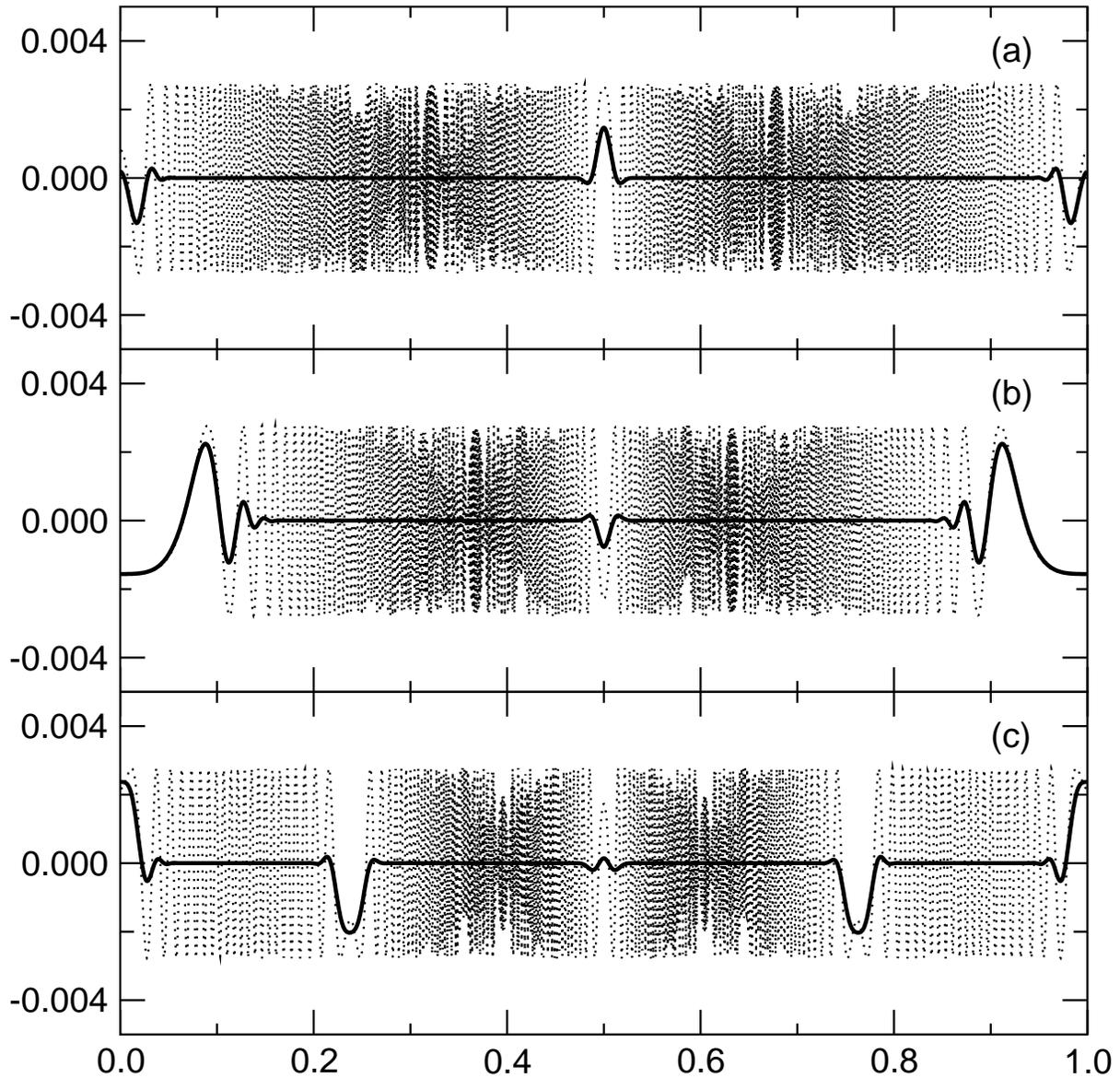}
\end{center}
\caption{$\sum_n|\Psi_{n}(Q)|^2\delta_{\varepsilon}(\theta-\theta_n)-1$ with
$\varepsilon = 2.2$, $\theta=\pi/4$ and $N=1597$ (dashed line) for the cosine perturbation. 
Also
 shown is a convolution of the data with a normalized Gaussian of width 0.007
(bold line). In (a) $\kappa=1.0$ and there is only one real unstable fixed 
point
(the central fixed point) at the origin; in
(b) $\kappa =2.0$, which is the parameter value at which the second order 
bifurcation takes place; in (c) $\kappa=3.0$, and there are
two new real
unstable fixed points at $q\approx 0.23$ and $q\approx 0.77$ 
in addition to the central fixed point, which is 
still located at the origin, but now is stable.}
\label{fig:f6}
\end{figure}

\begin{figure}[htb]
\begin{center}
\leavevmode
\psfig{figure=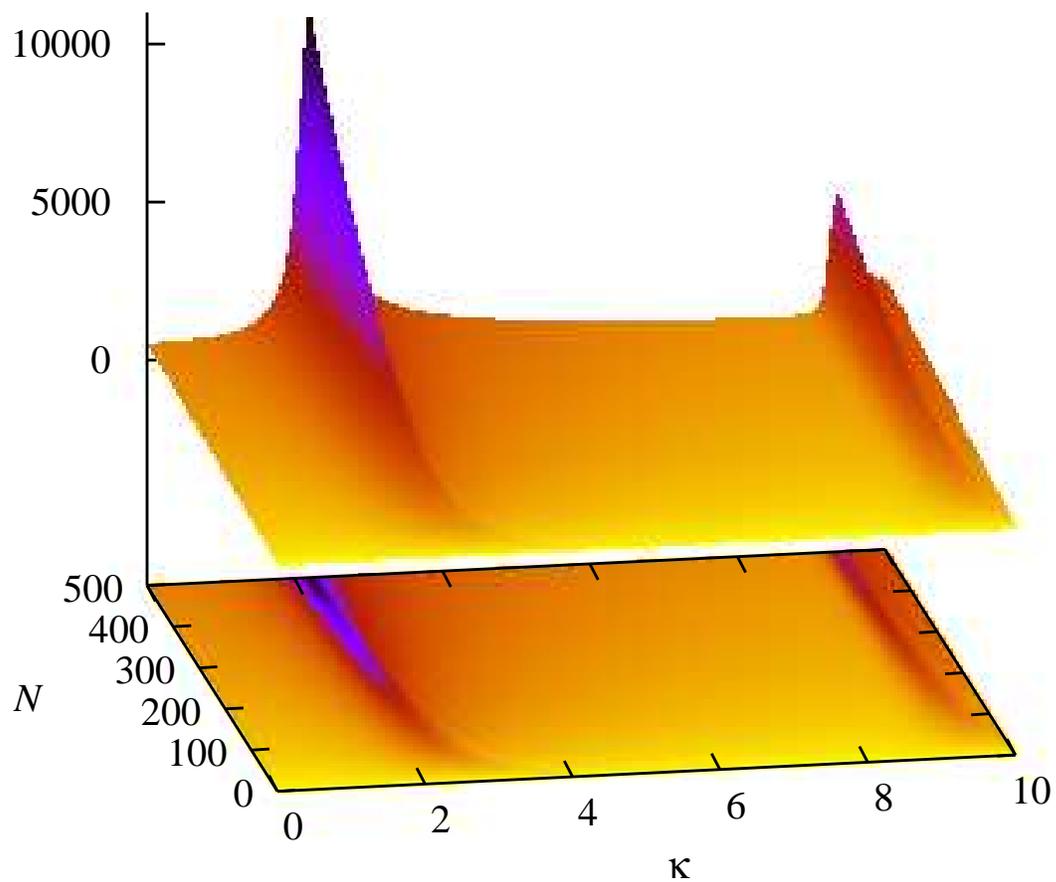,width=16.0cm,angle=0}
\end{center}
\caption{$|{\rm Tr}U(\kappa,N)|^2$ plotted as a function of $\kappa$ and $N$ for the cosine perturbation.  Note the two ridges: one 
due to the tangent bifurcation (at $\kappa \approx 9.208$) and 
the other due to the second order bifurcation (at $\kappa=2$).  
A projection of the data is shown underneath. }
\label{fig:f7}
\end{figure}

\begin{figure}[htb]
\begin{center}
\leavevmode
\psfig{figure=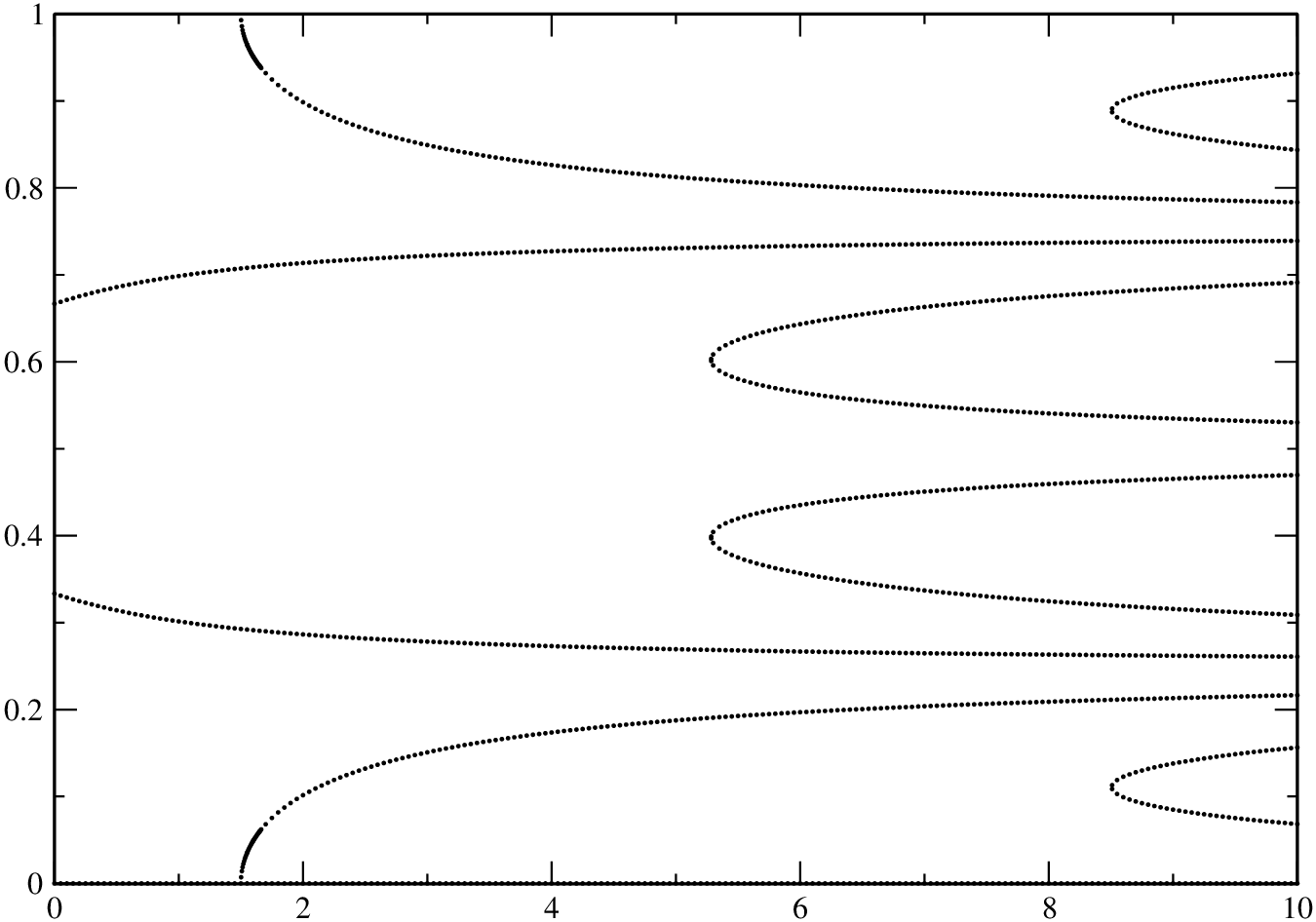,width=16.0cm,angle=0}
\end{center}
\caption{Predicted peak-centres in the 
autocorrelation function due to $l$-trajectories, computed by solving 
(\ref{f46a}) and (\ref{f46aa}), for the cosine perturbation.}
\label{fig:f8}
\end{figure}

\begin{figure}[htb]
\begin{center}
\leavevmode
\psfig{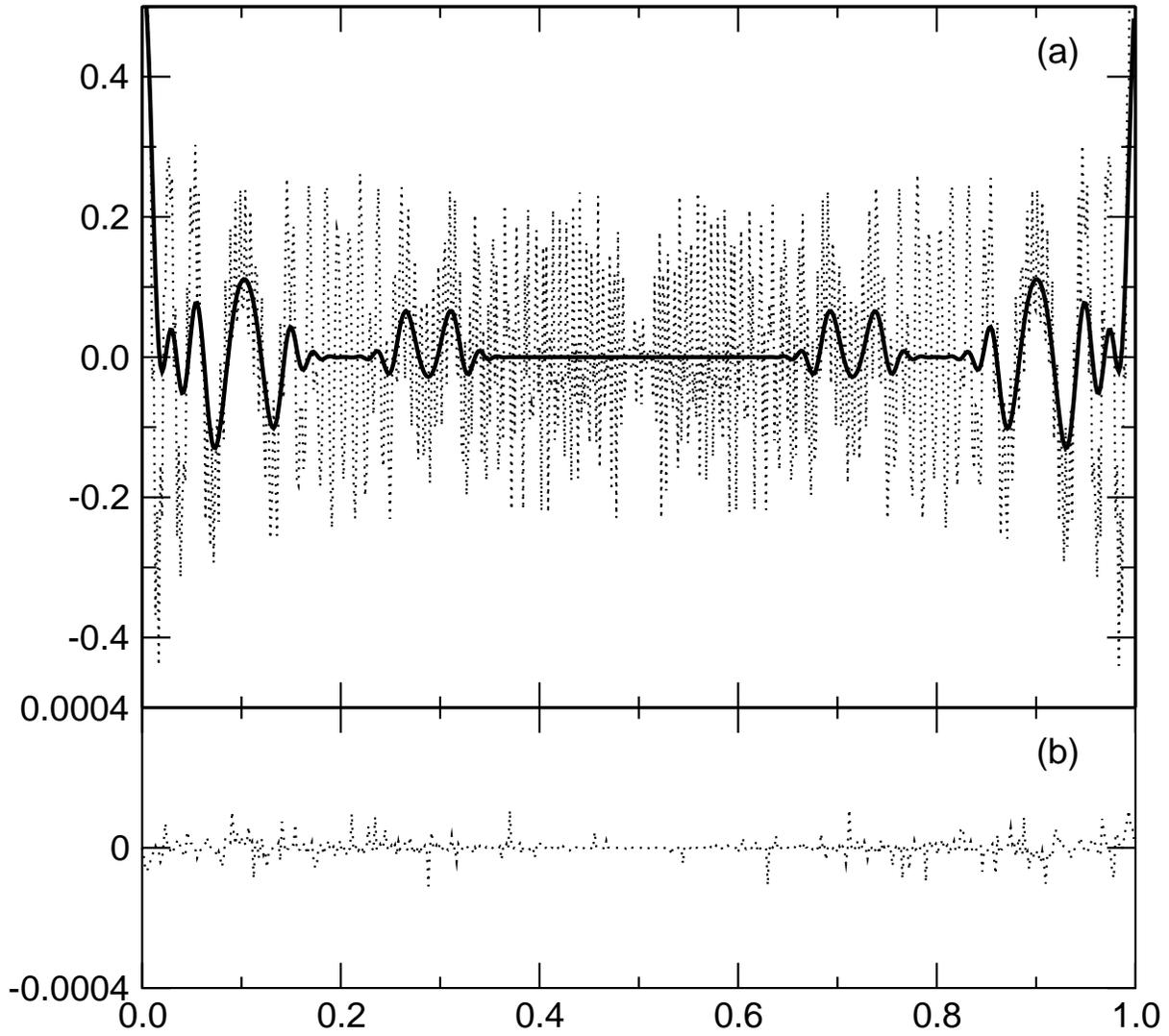}
\end{center}
\caption{(a) $c_r(L,\theta ,\epsilon)$ plotted as a function of $l=L/N$ 
for the cosine perturbation with 
$N=597$, $\kappa=2$,
 $\epsilon=1.5$ and $\theta=\pi/4$ (dashed line). The bold line  
represents a convolution of the data with a normalized Gaussian of
width 0.007. In (b) the difference between the exact
expression (\ref{f44are}) and the semiclassical approximation 
(\ref{f48}) is plotted.} 
\label{fig:f9}
\end{figure}

\begin{figure}[htb]
\begin{center}
\leavevmode
\psfig{figure=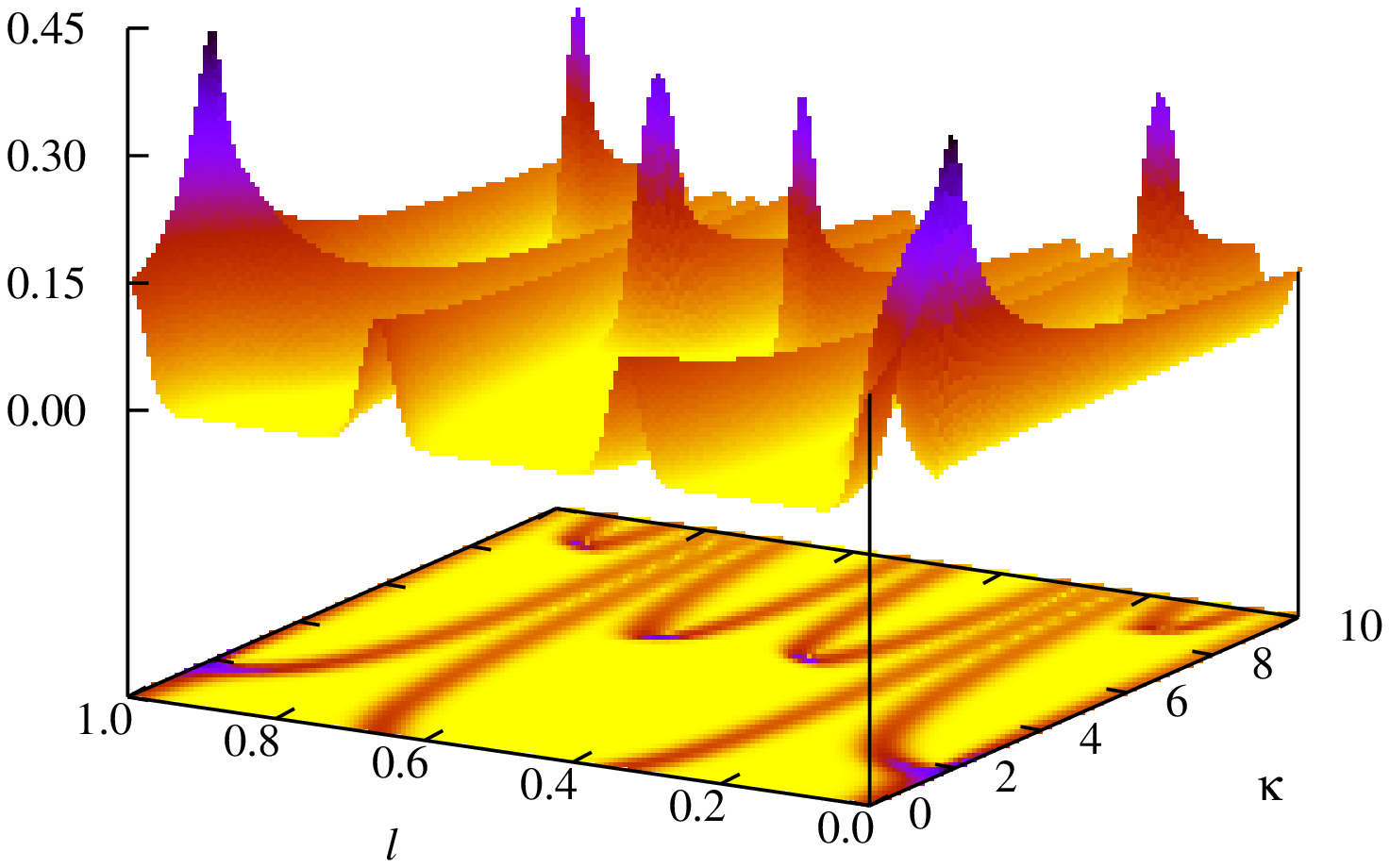,width=16.0cm,angle=0}
\end{center}
\caption{The absolute value of the convolution of the
 autocorrelation function (\ref{f44a}) with a normalized Gaussian  of width 0.01 plotted as a
function of $\kappa$ and $l=L/N$ for the cosine perturbation with $N=597$, $\epsilon=1.5$ and
$\theta=\pi/4$. 
A projection is shown underneath.  Compare the positions of the ridges with
the theoretical curves shown in Figure 8.  Note also the enhanced  
amplitude near the bifurcations (see also Figure 11) and the $l$-caustics.}
\label{fig:f10}
\end{figure}

\begin{figure}[htb]
\begin{center}
\leavevmode
\psfig{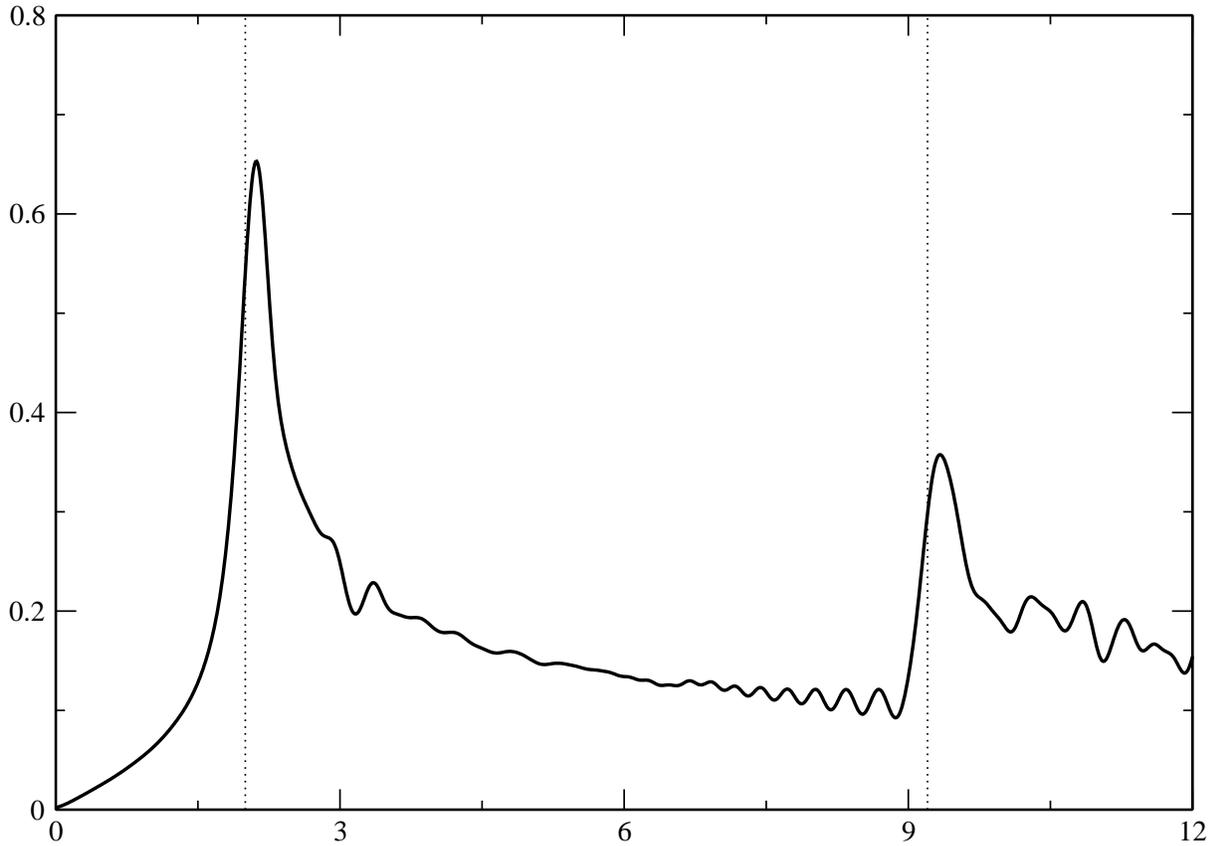}
\end{center}
\caption{A convolution of the absolute value of the
real part of the  autocorrelation function (\ref{f44are}) at $L=0$ with a normalized Gaussian in $\kappa$ of width 0.1, 
plotted as a
function of $\kappa$ for the cosine perturbation with $N=597$, 
$\epsilon=1.5$ and
$\theta=\pi/4$. 
Note the peaks associated 
with the tangent bifurcation (at $\kappa \approx 9.208$) and 
the second order bifurcation at ($\kappa=2$).}
\label{fig:f11}
\end{figure}

\end{document}